\documentclass[aps,pre,twocolumn,amsmath,amssymb,nofootinbib,eqsecnum,tighten,showpacs,floatfix]{revtex4-1}

\renewcommand{\vr}{\mathbf{r}}
\newcommand{\vR}{\mathbf{R}}

\usepackage{graphicx}
\usepackage{dcolumn} 
\usepackage{bm}      
\usepackage{epic}
\usepackage{color}

\begin{document}

\title{Semiclassical theory of speckle correlations}

\author{Maxim Breitkreiz} 
\affiliation{\mbox{Dahlem Center for Complex Quantum Systems and Fachbereich Physik, Freie Universit\"at Berlin, 14195 Berlin, Germany}}
\affiliation{\mbox{Institute of Theoretical Physics, Technische Universit\"at Dresden, 01062 Dresden, Germany}}
\altaffiliation[present address]{}
\author{Piet W.\ Brouwer} 
\affiliation{\mbox{Dahlem Center for Complex Quantum Systems and Fachbereich Physik, Freie Universit\"at Berlin, 14195 Berlin, Germany}}

\date{October 1, 2013}

\begin{abstract}
Coherent wave propagation in random media results in a characteristic 
speckle pattern, with spatial intensity correlations with short-range 
and long-range behavior. Here, we show how the speckle correlation 
function can be obtained from a ray picture for two representative
geometries: A chaotic cavity and a random waveguide. Our calculation allows
us to study the crossover between a ``ray limit'' and a ``wave
limit'', in which the Ehrenfest time $\tau_{\rm E}$ is larger
or smaller than the typical transmission time $\tau_{\rm D}$, 
respectively. 
Remarkably, long-range speckle correlations persist in the ray
limit $\tau_{\rm E} \gg \tau_{\rm D}$.
\end{abstract}


\maketitle

\section{Introduction}
Interference between multiply scattered waves causes large reproducable fluctuations of the intensity of radiation transmitted through a random medium or reflected off an irregular surface. These fluctuations, which are known as the ``speckle pattern'', can be seen as a fingerprint of the microscopic realization of the random medium. Speckle patterns have been observed for a wide variety of wave types, ranging from radio waves to optical light.

Speckle patterns of waves transmitted through random media have received particular interest in the last two decades, because they can be subjected to a statistical analysis and compared to theoretical predictions that involve only a few system-specific parameters \cite{shapiro1986,gwilich1987,feng1988}, such as the mean free path or the total transmission \cite{maret1995}. Quantitatively, speckle correlations are described with the help of the correlation function
\begin{equation}
  C(x-x') = \frac{\langle I(x) I(x') \rangle - \langle I(x) \rangle \langle I(x') \rangle}{\langle I(x) \rangle \langle I(x') \rangle},
  \label{eq:intro02}
\end{equation}
where $I$ is the intensity of the transmitted wave, $x-x'$ refers to a difference of a control parameter, such as the position of the source, the position of the detector, or the frequency, and the brackets $\langle \ldots \rangle$ indicate an average over a range of frequencies. In the theoretical and experimental literature, the correlation function $C$ is commonly written as the sum of three contributions \cite{gwilich1987,feng1988},
\begin{equation}
  C = C_1 + C_2 + C_3,
\end{equation}
which are consecutively smaller, but also decay slower upon increasing the difference between $x$ and $x'$. Whereas the original experiments could determine the short-range correlation function $C_1$ only \cite{etemad1986,freund1988,garcia1989}, later experiments were able to also identify the intermediate-range $C_2$ contribution \cite{vanalbada1990,deboer1992}, and even the long-range $C_3$ contribution \cite{sebbah2002}. 

In a random medium radiation generically changes its propagation direction multiple times between injection and detection. This can be either a smooth change, as in the case of a spatial gradient of the index of refraction, or an abrupt change, as for specular reflection off a mirror. In either case, in the limit that the wavelength $2 \pi/k$ is much smaller than the characteristic length scale of the ``scattering event'' (such as the length scale over which the index of refraction varies or the radius of curvature of a perfect mirror), the radiation can be considered to follow well-defined rays. The goal of this article is to present a theoretical investigation of the correlation function $C$ for the case that such a ``ray description'' applies to each individual scattering event in the random medium. This situation is of fundamental interest, since it elucidates the fate of the three contributions to the speckle correlations --- unambiguously a phenomenon belonging to the realm of wave optics ---, in the 
domain of (classical) ray optics. Although the conditions for a strict ray description are not met in most experiments listed above, either because the sizes of scatterers are small in comparison to the wavelength \cite{vanalbada1990,deboer1992}, or because the scattering occurs from only partially reflecting objects \cite{sebbah2002}, they can be met, {\em e.g.}, in experiments on microwave cavities, where reflection takes place off metal discs with a size that can be larger than the wavelength \cite{stoeckmann1990,sridhar1991,graef1992,stein1992}.

To explore how the (generically) chaotic classical dynamics in a random medium affects the speckle correlations it is instructive to consider the evolution of a wave packet passing through the random medium. In contrast to a classical point particle, which always follows a well-defined trajectory, wave packets naturally spread out, and the propagation of a wave packet can be mapped onto a single ray only as long as its size remains small enough. As soon as different parts of a wave packet start to evolve in an uncorrelated manner, the single-ray picture fails and a full \emph{wave} description is required. The characteristic time at which this crossover takes place is known as the ``Ehrenfest time'', 
\begin{equation}
  \tau_{\rm E} = \frac{1}{\lambda} \ln (k l),
  \label{eq:ehrenfesttime1}
\end{equation}
where $\lambda$ is the Lyapunov exponent of the ray trajectories in the random medium (which we assume to be chaotic) \cite{ott2002}, and $l$ a characteristic length scale of the ray dynamics, such that rays separated by a distance larger than $l$ must be considered uncorrelated. The Ehrenfest time is the time it takes for two rays, initially a wavelength apart, to diverge under the influence of the chaotic classical dynamics and reach a distance comparable to the characteristic scale $l$. We will use the term ``ray limit'' to refer to the case that the Ehrenfest time $\tau_{\rm E}$ exceeds the typical propagation time $\tau_{\rm D}$ for radiation transmitted through the random medium, whereas we reserve the term ``wave limit'' for the opposite case $\tau_{\rm E} \ll \tau_{\rm D}$.

The Ehrenfest time $\tau_{\rm E}$ was originally introduced by Larkin and Ovchinnikov in the context of a quasiclassical description of superconductivity \cite{larkin1968}. It plays an important role in the field of quantum chaos \cite{zaslavsky1981}, and in the phase coherent transport of electrons through ballistic mesoscopic conductors \cite{aleiner1996}, the latter application being of particular relevance to the present problem because of the formal analogy of the time-independent Schr\"odinger and Helmholtz equations. In the context of electronic transport, the ``ray limit'' is a ``classical limit'', in which the quantum mechanical propagation is replaced by propagation along classical trajectories. At zero temperature, the crossover to the classical limit takes place if $\tau_{\rm E}$ is comparable to the dwell time $\tau_{\rm D}$, the time electrons spend inside a mesoscopic conductor. In the regime $\tau_{\rm E} \gg \tau_{\rm D}$ that transport is essentially classical, it is found that some of the 
signatures of quantum transport, such as weak localization and shot noise, disappear \cite{aleiner1996,agam2000,adagideli2003,rahav2005,whitney2006}, whereas others, such as the universal conductance fluctuations, remain finite \cite{tworzydlo2004,jacquod2004,brouwer2006}. Making use of the same theoretical framework as used in the context of electronic transport, we will show that the short-range $C_1$ correlations are independent of $\tau_{\rm E}$, whereas the longer-range $C_2$ and $C_3$ correlations behave similar to shot noise and conductance fluctuations, respectively. In particular, the $C_2$ correlations {\em disappear} in the ray limit, whereas the $C_3$ correlations {\em remain finite}.

The existing calculations of the full speckle correlation function $C(x)$ make use of diagrammatic perturbation theory \cite{shapiro1986,gwilich1987,feng1988}. Since diagrammatic perturbation theory is built on the limit of weak, diffractive scatterers, it naturally describes the wave limit. For the crossover to the ray limit, we must follow a different theoretical approach. The method we follow here is the trajectory-based semiclassical approach, which was originally developed in the context of electronic transport through ballistic mesoscopic conductors \cite{baranger1993,argaman1993,richter2002,heusler2006,brouwer2006,brouwer2007}.

Our calculations are performed for the specific setup of transmission of scalar waves through a random medium, where we use the positions $\vr$ of the source and $\vR$ of the detector, both taken in a plane perpendicular to the axis of the waveguide, as well as the frequency $\omega$ of the light as control parameters. The same geometry and the same choice of control parameters was considered by Sebbah {\em et al.} \cite{sebbah2002}, who calculated the correlation functions $C(\Delta \vr,\Delta \vR,\Delta \omega)$ for the special case of a disordered waveguide from diagrammatic perturbation theory and found that the three contributions to the correlation function have a remarkably simple dependence on the spatial control parameters $\Delta \vr$ and $\Delta \vR$,
\begin{eqnarray}
  C_1 &=& A_1(\Delta \omega)
  F_d(|\Delta \vr|) F_d(|\Delta \vR|), \nonumber \\
  C_2 &=& \frac{1}{g} {A_2(\Delta \omega)}
  [F_d(|\Delta \vr|) + F_d(|\Delta \vR|)], \nonumber \\
  C_3 &=& \frac{1}{g^2} A_3(\Delta \omega),
  \label{eq:intro01}
  \label{eq:sebbahresult}
\end{eqnarray}
where $g$ is the dimensionless conductance of the waveguide, the $A_j$ are functions of the frequency shift $\Delta \omega$ only, and $F_d$ is a short-range function of its argument that depends on the dimensionality of the waveguide, 
\begin{equation}
  F_2(r) = J_0(k r),\ \ F_3(r) =  \frac{\sin kr}{kr}
\end{equation}
The main finding of this article is that the distinctive spatial dependence of Eqs.\ (\ref{eq:sebbahresult}) remains valid in the crossover to the ray limit, whereby only the functions $A_2$ and $A_3$ are modified. In particular, $A_2$ vanishes in the limiting case $\tau_{\rm E} \gg \tau_{\rm D}$, whereas $A_3$ remains finite. Our results will be derived for scalar waves in a two-dimensional system, for the case of a quasi-one-dimensional geometry, as in Ref.\ \onlinecite{sebbah2002}, as well as for the case of a chaotic cavity. Both geometries have been realized in microwave experiments, see, {\em e.g.}, Refs.\ \onlinecite{garcia1989,stoeckmann1990,sridhar1991}.

The remainder of the article is organized as follows: In Sec.\ \ref{sec:2} we express the speckle correlation function $C$ for scalar waves in terms of a multiple sum over classical rays propagating from source to detector, following the methods of trajectory-based semiclassics. Calculations of the three contributions $C_1$, $C_2$, and $C_3$ to the speckle correlation function for the special cases of a random waveguide and a chaotic cavity are then given in Secs.\ \ref{sec:3}, \ref{sec:4}, and \ref{sec:5}. We conclude in Sec.\ \ref{sec:6}.

\section{Semiclassical formalism}
\label{sec:2}

\begin{figure}[tb]
\includegraphics[width=\columnwidth]{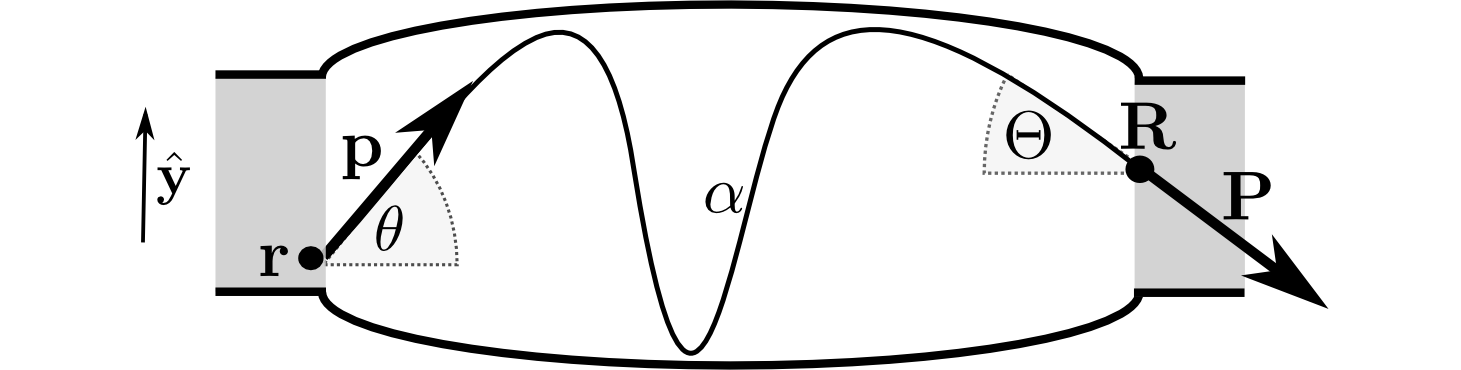}
\caption{\label{fig:1} Schematic picture of a trajectory $\alpha$ connecting the source at $\vr$ and the detector at $\vR$.}
\end{figure}

The precise geometry we consider is shown in Fig.\ \ref{fig:1}. It consists of a random medium connected to ideal waveguides on the left and the right. Waves at frequency $\omega$ originate from a source at position $\vr$ in the left waveguide and their intensity is detected at a detector at position $\vR$ in the right waveguide. The positions of source and detector can be varied in the direction  $\hat{\mathbf{y}}$ perpendicular to the axis of the waveguide, and the intensity fluctuations are measured as a function of $\omega$ and the component $y$ ($Y$) of the source (detector) position.

Inside the random medium, scattering takes place from perfect specularly reflecting mirrors with a radius of curvature that is large in comparison to the wavelength $2 \pi/k$. This condition ensures that the waves propagate along well-defined rays, which is the motivation of the ``ray limit'' taken in the calculation below. For simplicity, we will consider scalar waves in a two-dimensional geometry. This simplification is appropriate for quasi-two-dimensional microwave cavities, where microwaves have a unique polarization direction \cite{stoeckmann1990}.

Starting point of our calculation is an expression for the intensity $I_{\omega}(\vR,\vr)$ at the detector position $\vR$ in terms of the exact Green function $G_{\omega}^{\pm}(\vR,\vr)$ for propagation in the combined system consisting of the random medium and the waveguides \cite{feng1988},
\begin{equation}
  I_{\omega}(\vR,\vr) = G^{+}_{\omega}(\vR,\vr) G^{-}_{\omega}(\vR,\vr),
\end{equation}
where the Green functions are solutions of the time-independent Helmholtz equation
\begin{equation}
  (k^2 + \nabla^2 \pm i \eta) G^{\pm}_{\omega}(\vr',\vr) = \delta(\vr'-\vr),
\end{equation}
with $\omega = c k$, $c$ being the wave velocity, with $\eta$ a positive infinitesimal, and with the appropriate boundary conditions at the sample boundaries and mirrors.

The ray limit we are interested in corresponds to the limit of small wavelength $2 \pi/k$. In order to formally take this limit, and motivated by the trajectory-based semiclassical theory of electronic transport, we introduce a fictitious ``Planck's constant'' $\hbar$ by writing $k = p/\hbar$, where $p$ has the dimension of momentum. The product $p l$, with $l$ a characteristic length scale for the random medium, then has the dimension of action, and the ray limit corresponds to the limit $\hbar \to 0$, while keeping the typical action $p l$ fixed.

In this limit, the Green function $G_{\omega}^{\pm}(\vR,\vr)$ can be written as a sum over contributions from rays or ``trajectories'' $\alpha$ connecting the positions $\vr$ and $\vR$ \cite{gutzwiller1990,baranger1993,baranger1993b},
\begin{eqnarray}
  G^{+}_{\omega}(\vR,\vr) &=& G^{-}_{\omega}(\vR,\vr)^* \nonumber \\ &=&
  \sum_{\alpha:\vr \to \vR}
  \sqrt{\frac{i \hbar D_{\alpha}}{8 \pi p^2}}
  e^{\frac{i}{\hbar} S_{\alpha}(\omega)-i\frac{\pi}{2}\mu_{\alpha}},
  \label{eq:G}
\end{eqnarray}
where $S_{\alpha}(\omega)$ is the ``classical action'' of the ray $\alpha$,
\begin{equation}
  S_{\alpha}(\omega) = \int_{\alpha:\vr \to \vR} d\mathbf{l} \cdot
  \mathbf{p},
\label{eq:S}
\end{equation}
the Maslov-Morse index $\mu_{\alpha}$ gives the number of reflections at mirrors or sample boundaries \cite{baranger1993b}, and $D_{\alpha}$ is the so-called stability amplitude of the trajectory $\alpha$, the probability flux at $\vR$ along $\alpha$ for an isotropic source at $\vr$. Defining coordinates as in Fig.\ \ref{fig:1}, one has
\begin{eqnarray}
  D_{\alpha} &=&
  \frac{1}{\cos \theta \cos \Theta}
  \left| \frac{\partial P_y}{\partial y} \right| \nonumber \\ &=&
  \frac{1}{\cos \theta \cos \Theta}
  \left| \frac{\partial p_y}{\partial Y} \right|,
\end{eqnarray}
where one has $\cos \theta = ({1-p_y^2/p^2})^{1/2}$ and $\cos \Theta = ({1 - P_y^2/p^2})^{1/2}$.
For the system we consider, the momentum $p$ is a constant, so that the classical action $S_{\alpha}$ is directly proportional to the duration $\tau_{\alpha}$ of the ray $\alpha$,
\begin{equation}
  S_{\alpha}(\omega) = p\, \tau_{\alpha} c,
\end{equation}
with $c$ the wave velocity.

\begin{figure}[tb]
\includegraphics[width=\columnwidth]{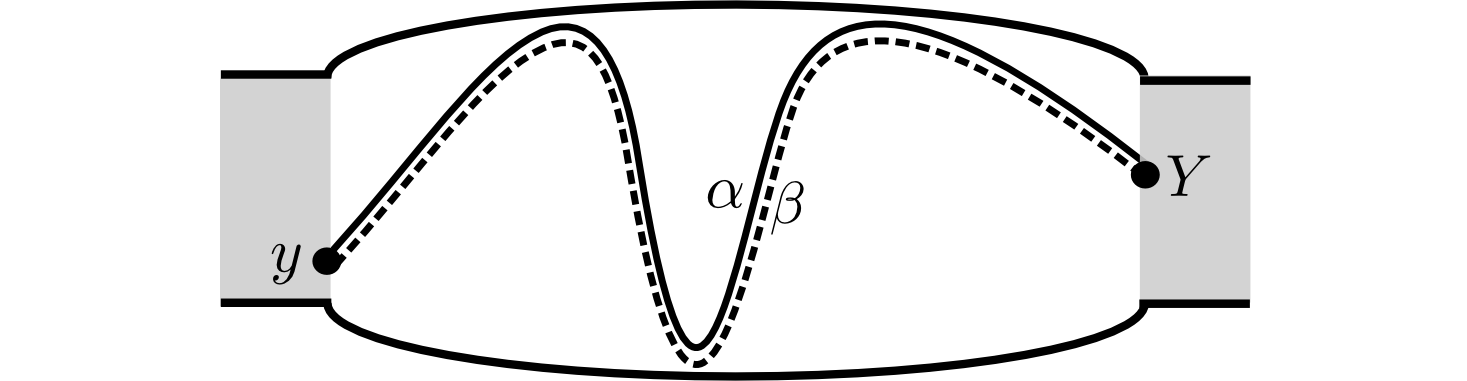}
\caption{Schematic picture of a diagonal trajectory contribution, $\alpha =\beta$, to the mean intensity $\langle I_{\omega}(Y,y) \rangle$.\label{fig:2}}
\end{figure}

Using the semiclassical expression for the Green function, the intensity $I_{\omega}(\vR,\vr)$ is written as a double sum over pairs of classical rays $\alpha$ and $\beta$ that connect the source at $\vr$ to the detector at $\vR$,
\begin{eqnarray}
  I_{\omega}(\vR,\vr) &=&
  \frac{\hbar}{8 \pi p^2}
  \sum_{\alpha,\beta}
  \sqrt{D_{\alpha} D_{\beta}} \nonumber \\ && \mbox{} \times
  e^{\frac{i}{\hbar}(S_{\alpha}(\omega) - S_{\beta}(\omega)) -i \frac{\pi}{2}(\mu_{\alpha} - \mu_{\beta})}, 
  \label{eq:I}
\end{eqnarray}
Upon performing the frequency average $\langle \ldots \rangle$, the phase factor $e^{(i/\hbar)(S_{\alpha} - S_{\beta})}$ in Eq.\ (\ref{eq:I}) has fast fluctuations, which will cancel the contributions of all terms in the summation (\ref{eq:I}), except for those for which there is a systematic correlation between the actions of the trajectories $\alpha$ and $\beta$. The leading configuration of such systematically correlated trajectories is the case $\alpha = \beta$ (see Fig.\ \ref{fig:2}), which gives
\begin{equation}
  \langle I(\vR,\vr) \rangle =
  \frac{\hbar}{8 \pi p^2}
  \sum_{\alpha} D_{\alpha}.
\end{equation}

It remains to perform the summation over trajectories. Hereto, we take a fixed cross section in the waveguides connecting to the system at the source and detector sides, and use the coordinate pair $(y,p_y)$ with $-p < p_y < p$ to parameterize rays entering the system from the source side at position $y$ and with momentum $p_y$ perpendicular to the waveguide axis, and the coordinate pair $(Y,P_y)$ with $-p < P_y < p$ to parameterize rays exiting the system at the detector side. For a trajectory that is transmitted from the source side to the detector side and that enters the system at coordinates $(y,p_y)$, we define $y_{{\rm out}}(y,p_y)$ and $p_{y,{\rm out}}(y,p_y)$ as the coordinates upon exit. One then has
\begin{eqnarray}
  \sum_{\alpha} D_{\alpha} &=&
  \int_{-p}^{p} dp_y dP_y \\ && \nonumber \mbox{} \times
  \frac{\delta(y_{{\rm out}}(y,p_y) - Y) \delta(p_{y,{\rm out}}(y,p_y) - P_y)}
  {\cos \theta \cos \Theta}.
\end{eqnarray}
Upon taking into account small fluctuations around $y$ or $Y$, one may replace the product of delta functions by the probability density $p(Y,P_y;y,p_y)$ that a ray entering on the source side at $(y,p_y)$ exits at the detector side at $(Y,P_y)$, which gives
\begin{equation}
  \langle I_{\omega}(Y,y) \rangle =
  \frac{\hbar}{8 \pi p^2}
  \int_{-p}^{p} dp_y dP_y
  \frac{p(Y,P_y;y,p_y)}{\cos \theta \cos \Theta},
\end{equation}
where we wrote $I_{\omega}(Y,y)$ instead of $I_{\omega}(\vR,\vr)$ in order to make manifest that we only consider variations of the position of source and detectors in a plane perpendicular to the waveguide axes. We have
\begin{equation}
  p(Y,P_y.;y,p_y) = \frac{T}{2 W p},
\end{equation}
where $T$ is the transmission probability, which is different for the case of a chaotic cavity (CC) and a quasi-one-dimensional random waveguide (WG). One has $T = 1/2$ for a chaotic cavity and $T = l/L$ for a random waveguide of length $L$ and transport mean free path $l$. The transmission probability is related to the ``dimensionless conductance'' $g$,
\begin{equation}
  g = \frac{k W}{\pi} T.
\end{equation}
Restoring $p = \hbar k$, we find that the average intensity is given by
\begin{eqnarray}
  \langle I_{\omega}(Y,y)\rangle &=&
  \frac{\pi^2 g}{16 k^2 W^2} \nonumber \\ &=&
  \frac{\pi}{16 k W} \times 
  \begin{cases} 1/2 & (\mbox{CC}), \\
  l/L & (\mbox{WG}). \end{cases}
  \label{eq:Iavg}
\end{eqnarray}

\begin{figure}[tb]
\includegraphics[width=\columnwidth]{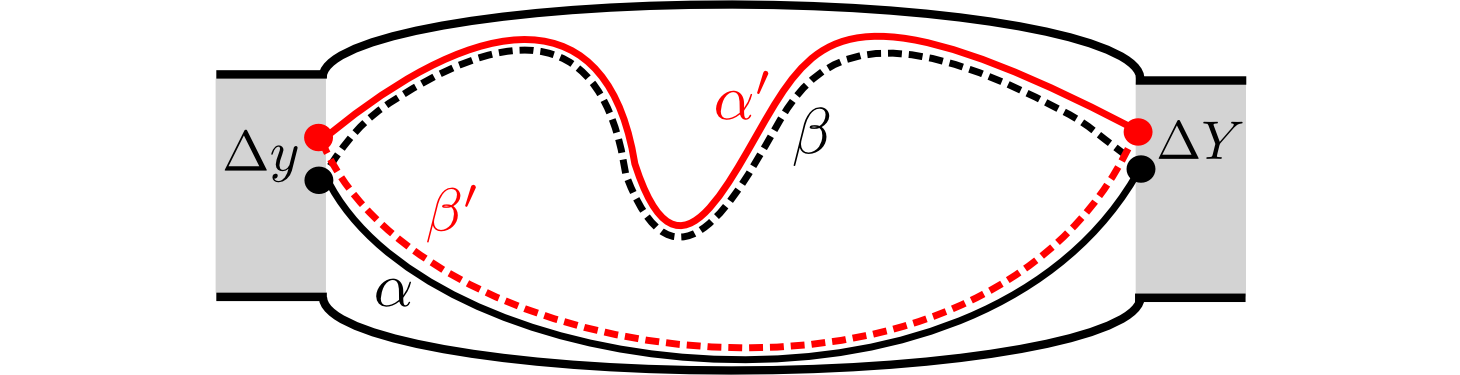}
\caption{(color online) Trajectory constellation contributing to the $C_1$ contribution of the speckle correlation function.\label{0-encb}}
\end{figure}

In a similar way, the product of two intensities that enters into the correlation function $C$ becomes a double sum over pairs of classical rays $\alpha$ and $\beta$ that connect the source at $y$ to the detector at $Y$ and pairs of rays $\alpha'$ and $\beta'$ that connect the source at $y'$ to the detector at $Y'$,
\begin{eqnarray}
\label{eq:II}
  \lefteqn{I_{\omega}(Y,y)\, I_{\omega'}(Y',y')} 
  \nonumber \\ &=&
  \left( \frac{\hbar}{8 \pi p^2} \right)^2 \sum_{\alpha,\beta}\,
\sum_{\alpha',\beta'}
\sqrt{D_{\alpha}D_{\alpha'}D_{\beta}D_{\beta'}}\, e^{\frac{i}{\hbar}\Delta S}
  \nonumber \\ && \mbox{} \times
  e^{-i\frac{\pi}{2}(\mu_{\alpha} - \mu_{\beta} + \mu_{\alpha'} - \mu_{\beta'})}
,\nonumber 
\end{eqnarray}
with the action difference
\begin{equation}
\Delta S=S_{\alpha}(\omega)-S_{\beta}(\omega)+S_{\alpha'}(\omega')-S_{\beta'}(\omega') 
\label{eq:DS}
\end{equation}
Again, upon performing the average $\langle \ldots \rangle$ the phase $\Delta S/\hbar$ in the exponent in Eq.\ (\ref{eq:II}) has fast fluctuations, and only trajectory configurations for which there is a systematic correlation between the four trajectories $\alpha$, $\beta$, $\alpha'$, and $\beta'$, such that the action difference $\Delta S$ is small, will contribute to the sum in Eq.\ (\ref{eq:II}). The leading configuration of such systematically correlated trajectories is the case $\alpha = \beta$, $\alpha' = \beta'$. This diagonal contribution to the correlation function factorizes and cancels precisely against the product of separately averaged intensities $\langle I_{\omega}(Y,y)\rangle \langle I_{\omega'}(Y',y') \rangle$, for which each factor is given by Eq.\ (\ref{eq:Iavg}) above. To leading order in the system's dimensionless conductance $g$, one finds three other trajectory configurations that contribute to $C$, which give rise to the three contributions $C_1$, $C_2$, and $C_3$ to the correlation 
function. These three contributions will be discussed separately in the next three Sections.

\section{$C_{1}$ contribution}
\label{sec:3}

The first contribution $C_{1}$ is the largest contribution to the correlation function and thus describes the most visible part of the speckle pattern --- the large intensity fluctuations. The trajectories that contribute to the $C_{1}$ contribution are shown schematically in Fig.\ \ref{0-encb} \cite{berkovits1989}: The trajectory $\alpha$ is paired with $\beta'$, and $\alpha'$ is paired with $\beta$. This pairing is possible only if the distances $\Delta y$ and $\Delta Y$ between the source positions and the detector positions are small, which leads to the double short range behavior of $C_{1}$ correlations. The trajectory configuration of Fig.\ \ref{0-encb} bears a close resemblance to the diagrams corresponding to the $C_{1}$ contribution in diagrammatic perturbation theory \cite{shapiro1986,akkermans2010}.

We note that the choice of the trajectories $\alpha$ and $\alpha'$ uniquely fixes the remaining two trajectories $\beta$ and $\beta'$. The differences $\Delta y$ and $\Delta Y$ must be of order of the wavelength $2\pi/k$ to ensure that the action difference $\Delta S$ is of order $\hbar$.  The trajectories $\alpha$ and $\beta'$ have the same angles $\theta$ and $\Theta$ with the waveguide axis at source and detector, respectively, up to an unimportant difference of order $\hbar/(p l) \ll 1$, which we neglect. The same holds for the trajectories $\alpha'$ and $\beta$, for which the angles with the waveguide at source and detector are denoted $\theta'$ and $\Theta'$, respectively. This allows a straightforward calculation of the action difference $\Delta S$, which can be written as a sum of contributions related to the trajectory configuration at the source, at the detector, and related to the frequency difference $\Delta \omega$,
\begin{eqnarray}
\Delta S &=& \Delta S_{\rm source} + \Delta S_{\rm detector} + \Delta S_{\omega},
  \label{eq:dSC1}
\end{eqnarray}
with
\begin{eqnarray}
  \Delta S_{\rm source} &=& \Delta y(p_y' - p_y), \nonumber \\ 
  \Delta S_{\rm detector} &=& \Delta Y(P_y' - P_y), \nonumber \\
  \Delta S_{\omega} &=& c \Delta p (\tau_{\alpha} - \tau_{\alpha'}),
  \label{eq:dS1}
\end{eqnarray}
with $\Delta p = \hbar \Delta \omega/c$.
For the stability amplitudes we may set $D_{\beta} = D_{\alpha'}$ and $D_{\beta'} = D_{\alpha}$, up to corrections of relative size $\hbar/(p l)$, so that 
\begin{eqnarray}
  C_1 &=&
  \left( \frac{\hbar}{8 \pi p^2 \langle I \rangle} \right)^2
  \nonumber \\ && \mbox{} \times
  \sum_{\alpha,\alpha'} D_{\alpha} D_{\alpha'}
  e^{i \Delta S/\hbar},
\end{eqnarray}
with $\Delta S$ given by Eq.\ (\ref{eq:dSC1}).
Replacing the summation over trajectories by an integral over probabilities as in the derivation of Eq.\ (\ref{eq:Iavg}), we find
\begin{eqnarray}
  C_1 &=&
  \left( \frac{\hbar}{8 \pi p^2 \langle I \rangle} \right)^2
  \int_{-p}^{p} dp_y dp_y' dP_y dP_y'
  \int_0^{\infty} d \tau d\tau' \nonumber \\ && \mbox{} \times
  \frac{p(Y,P_y;y,p_y;\tau) p(Y,P_y';y,p_y';\tau')}{\cos \theta
  \cos \theta' \cos \Theta \cos \Theta'}
  e^{i \Delta S/\hbar},
\end{eqnarray}
where now $p(Y,P_y;y,p_y;\tau)$ is the probability density that a trajectory entering at the source side at coordinates $(y,p_y)$ exits on the detector side at coordinates $(Y,P_y)$ after a propagation time $\tau$ inside the system. For the two geometries we consider, we have
\begin{equation}
  p(Y,P_y;y,p_y; \tau) =
  \frac{T}{2 W p}
  p_{\tau}(\tau),
  \label{eq:ptrans}
\end{equation}
where $p_{\tau}(\tau)$ is the ``dwell time distribution'' for transmitted trajectories, which is
\begin{equation}
  p_{\tau}(\tau) = \frac{1}{\tau_{\rm D}} e^{-\tau/\tau_{\rm D}}
\end{equation}
for a chaotic cavity and
\begin{equation}
  p_{\tau}(\tau) = \frac{8}{\pi^{2}\tau_{D}^{2}}\tau 
  \sum_{n=0}^{\infty} (2n+1)^{2}e^{-\frac{(2n+1)^{2}\tau}{\tau_{D}}},  
\end{equation}
for a quasi-one-dimensional random waveguide with diffusion coefficient $D = lc/\pi$. In both cases, $\tau_{\rm D}$ is a characteristic classical dwell time, which is $\tau_{\rm D} = \pi A/2 W c$ for a chaotic cavity of area $A$, and $\tau_{\rm D} = L^2/D \pi^2$ for the random waveguide.

Substituting Eq.\ (\ref{eq:ptrans}) for the probabilities $p(Y,P_y;y,p_y;\tau)$ and $p(Y,P_y';y,p_y';\tau')$, the integral factorizes and one finds
\begin{equation}
  C_1(\Delta y,\Delta Y,\Delta \omega) =
  J_0(k \Delta y)^2 J_0(k \Delta Y)^2
  f(\Delta \omega),
\end{equation}
with $J_0$ the Bessel function of the first kind and 
\begin{equation}
  f(\Delta \omega) =
  \frac{1}{1 + (\tau_{\rm D}\Delta \omega )^2} 
\end{equation}
for the case of a chaotic cavity and 
\begin{equation}
  f(\Delta \omega) = \frac{\pi^2 \tau_{\rm D} \Delta \omega}
  {|\sinh(\pi \sqrt{i \tau_{\rm D} \Delta \omega})|^2}
\end{equation}
for the random waveguide.
One verifies that $C_1 \to 1$ in the simultaneous limit $\Delta y$, $\Delta Y$, and $\Delta \omega \to 0$, consistent with the Poisson statistics of the intensity at a single position and frequency.

\section{$C_{2}$ contribution}
\label{sec:4}

\begin{figure}[tb]
\includegraphics[width=\columnwidth]{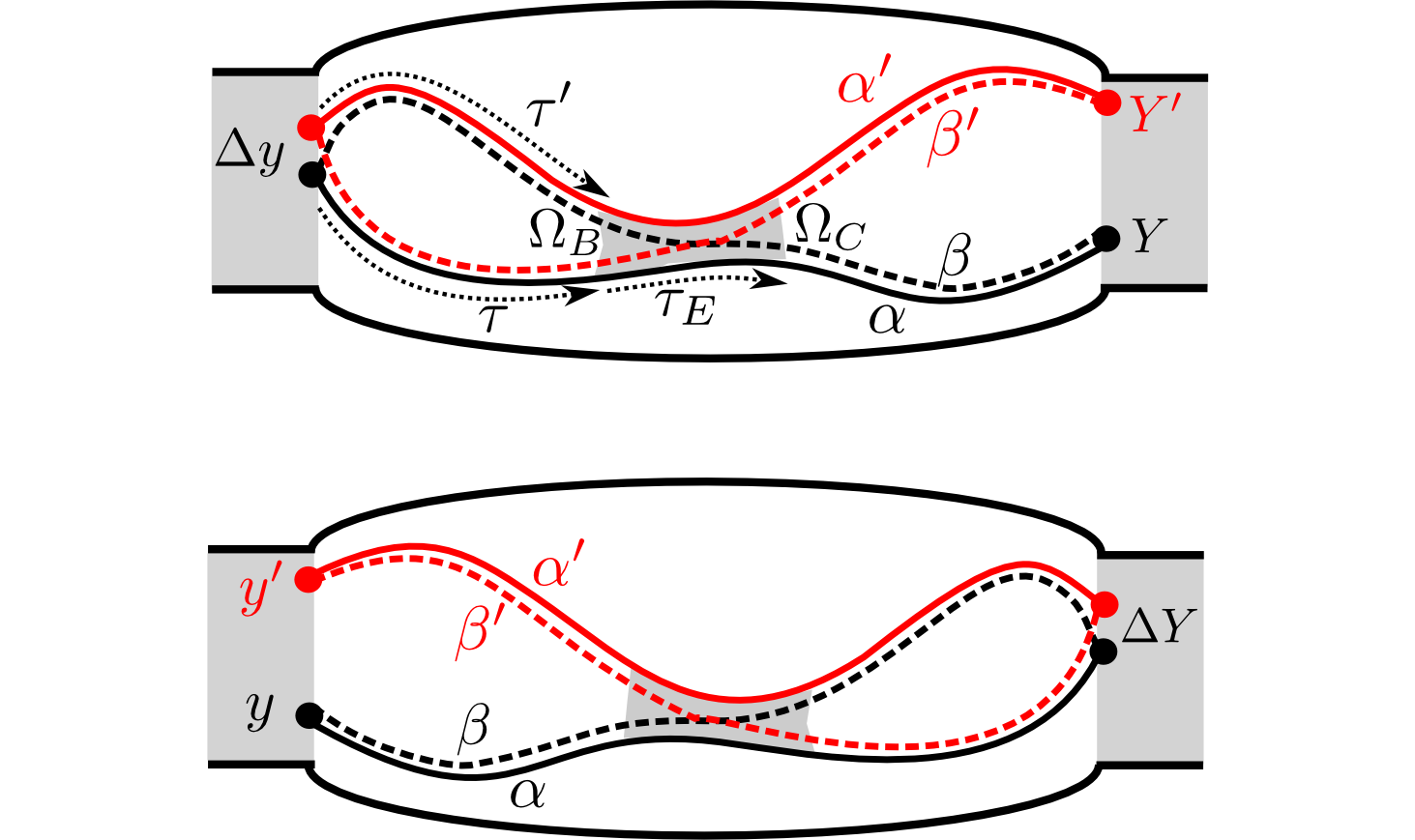}
\caption{(color online) Trajectory constellation contributing to $C_2$. The top panel shows the trajectory configurations for which correlations remain short-ranged as a function of the source position difference $\Delta y$; the bottom panel refers to short-ranged correlation as a function of the detector position difference $\Delta Y$. In both panels, the trajectories undergo a small-angle encounter, indicated by the grey area. Phase space points at the beginning and end of the encounter are denoted $\Omega_B$ and $\Omega_C$, respectively.\label{1-enc}}
\end{figure}

The trajectory constellations which give the leading contribution to the correlation function when either the sources or the detectors (but not both) are far apart are shown schematically in Fig.\ \ref{1-enc}. As in the case of the $C_1$ contribution of the previous section, these trajectories bear close resemblance to the corresponding diagrams in the diagrammatic calculation. As shown in the figure, there are two such constellations, one in which the detectors are apart, but the sources are not (upper panel in Fig.\ \ref{1-enc}) and one in which the sources are apart, but the detectors are not (lower panel). Below we focus on the former contribution, which we denote $C_{2,1}$.

The trajectory constellation of the upper panel of Fig.\ \ref{1-enc} shows a small-angle encounter of the four trajectories involved, such that the trajectories $\alpha$ and $\beta'$, as well as $\alpha'$ and $\beta$ are paired on the source-side of the encounter, whereas $\alpha$ is paired with $\beta$ and $\alpha'$ with $\beta'$ on the detector side of the encounter. Similar trajectory configurations occur in the semiclassical calculation of the shot noise power in electronic transport. Of particular relevance to the present calculation is Ref.\ \onlinecite{brouwer2007b}, where the shot noise power was calculated semiclassically for the two geometries we are interested in (See also Refs.\ \onlinecite{whitney2006,mueller2007,brouwer2006c} for related calculations involving a chaotic cavity only). Below we will adapt the calculation of Ref.\ \onlinecite{brouwer2007b} to the $C_2$ contribution to the speckle correlation function.

As in the case of the $C_1$ contribution, we note that the two trajectories $\alpha$ and $\alpha'$ uniquely fix the remaining two trajectories $\beta$ and $\beta'$. For the trajectory configuration of the upper panel of Fig.\ \ref{1-enc} there are three contributions to the action difference,
\begin{equation}
  \Delta S = \Delta S_{\rm source} + \Delta S_{\omega} + \Delta S_{\rm enc},
  \label{eq:dSC2}
\end{equation}
where the contribution $\Delta S_{\rm source}$ related to the distance $\Delta y$ between the sources is given in Eq.\ (\ref{eq:dS1}) of the previous section, the contribution $\Delta S_{\omega}$ related to the frequency difference $\Delta \omega$ is
\begin{equation}
  \Delta S_{\omega} = c (\tau - \tau') \Delta p,
\end{equation}
where $\tau$ and $\tau'$ are the propagation times along $\alpha$ and $\alpha'$ from the source to the beginning of the encounter, respectively (see Fig.\ \ref{1-enc}), and the countribution $\Delta S_{\rm enc}$ is related to the small-angle encounter between the trajectories $\alpha$ and $\alpha'$. Finally, since the trajectories are pairwise equal throughout, one has
\begin{equation}
  \sqrt{D_{\alpha} D_{\alpha'} D_{\beta} D_{\beta'}} = D_{\alpha} D_{\alpha'},
\end{equation}
so that 
\begin{equation}
  C_{2,1} = \left( \frac{\hbar}{8 \pi p^2 \langle I \rangle} \right)^2
  \sum_{\alpha,\alpha'} D_{\alpha} D_{\alpha'}
  e^{i \Delta S/\hbar},
\end{equation}
with $\Delta S$ given by Eq.\ (\ref{eq:dSC2}).

In order to parameterize the trajectories $\alpha$ and $\alpha'$ we not only need the coordinates specifying the entrance and exit from the system, but also the phase space coordinates $\Omega_{B}$ and $\Omega_{C}$ specifying the beginning and end of the encounter (see Fig.\ \ref{1-enc}). Locally, around a reference trajectory, the phase space coordinates $\Omega$ consist of the transverse momentum $p_{\perp}$, the transverse distance $r_{\perp}$, and the propagation time $t$ along the trajectory. The beginning and end of the encounter are defined as those points along the trajectories, where the phase space distance between $\alpha$ and $\alpha'$ is large enough, that the propagation of the two trajectories can be considered uncorrelated. Typically, the encounter ends when $|\Delta p_{\perp}| \sim p$ or $|\Delta r_{\perp}| \sim l$, whichever occurs first.

The encounter-related contribution $\Delta S_{\rm enc}$ has been calculated in Refs.\ \onlinecite{sieber2001,spehner2003}. As shown in Ref.\ \onlinecite{sieber2001}, only encounters of a duration $\tau_{\rm E}$ contribute to the trajectory sum, $\tau_{\rm E}$ being the Ehrenfest time defined in Eq.\ (\ref{eq:ehrenfesttime1}). The inclusion of the action difference $\Delta S_{\rm enc}$ and the corresponding summation over trajectories proceeds completely analogous to the calculation of the shot noise power. Referring to Ref.\ \onlinecite{brouwer2007b} for details of this part of the calculation, the summation over trajectories $\alpha$ and $\alpha'$ can then be written in terms of a double integration over the phase space points $\Omega_B$ and $\Omega_C$ for the beginning and end of the encounter, 
\begin{eqnarray}
  C_{2,1} &=& 2 \pi \hbar 
  \left( \frac{\hbar}{8 \pi p^2 \langle I \rangle} \right)^2
  \int_{-p}^{p} dp_y dp_y' dP_y dP_y'
  \int_0^{\infty} d\tau d\tau' \nonumber \\ && \mbox{} \times
  \int d\Omega_{B} d\Omega_{C}
  \frac{p_{\rm s}(\bar{\Omega}_{B};y,p_y;\tau) 
  p_{\rm s}(\bar{\Omega}_{B};y,p_y';\tau')}
  {\cos \theta \cos \theta'}
  \nonumber \\ && \mbox{} \times
  \frac{p_{\rm d}(\Omega_{C};Y,P_y) p_{\rm d}(\Omega_{C};Y',P_y')}
  {\cos \Theta \cos \Theta'}
  \nonumber \\ && \mbox{} \times
  \frac{\partial}{\partial \tau_{\rm E}}
  p(\Omega_{C},\Omega_{B};\tau_{\rm E})
  e^{i (\Delta S_{\rm source} + \Delta S_{\omega})/\hbar},
\end{eqnarray}
In this expression the phase space volume element $d\Omega = dp_{\perp} dr_{\perp} dt$ and $\bar \Omega$ denotes the time-reversed of the phase space point $\Omega$. Further, $p_{\rm s}(\bar{\Omega}_B;y,p_y;\tau)$ is the probability density that a ray starting at phase space point $\bar{\Omega}_B$ exits the system at the source side at coordinate $(y,p_y)$ and after a propagation time $\tau$. Similarly, $p_{\rm d}(\Omega_C;Y,P_y)$ is the probability density that a ray starting at phase space point $\Omega_C$ exits the system at the detector side at coordinates $(Y,P_y)$ (irrespective of propagation length). Finally, $p(\Omega_C,\Omega_B,\tau)$ is the phase space probability density that a ray starting at phase space point $\Omega_B$ is found at phase space point $\Omega_C$ after a propagation time $\tau$. In order to perform the integrations over the momenta at source and detector we make use of the relations
\begin{equation}
  p_{\rm s}(\bar{\Omega}_B;y,p_y;\tau) =
  \frac{1}{2 p W} p_{\rm s}(\bar{\Omega}_B;\tau)
\end{equation}
and
\begin{equation}
  p_{\rm d}(\Omega_C;Y,P_y) =
  \frac{1}{2 p W} p_{\rm d}(\Omega_C),
\end{equation}
where $p_{\rm s}(\Omega_B;\tau)$ is the probability density that a ray starting at $\Omega_B$ exits the system at the source side after propagation time $\tau$ and $p_{\rm d}(\Omega_C)$ is the probability density that a ray starting at $\Omega_C$ exits the system at the drain side. For a chaotic cavity, one has
\begin{equation}
  p_{\rm s}(\Omega;\tau) =
  \frac{1}{2 \tau_{\rm D}} e^{-\tau/\tau_{\rm D}},\ \
  p_{\rm d}(\Omega) = \frac{1}{2},
  \label{eq:pchaotic1}
\end{equation}
whereas for a one-dimensional random waveguide one has
\begin{eqnarray}
  \label{eq:wg1}
  p_{\rm s}(\Omega;\tau)
  &=& \sum_{n=1}^{\infty} \frac{2 n}{\pi \tau_{\rm D}}
  \sin \frac{n \pi x_{\Omega}}{L} e^{-n^2 \tau/\tau_{\rm D}},
  \nonumber \\
  p_{\rm d}(\Omega) &=& \frac{x_{\Omega}}{L},
\end{eqnarray}
with $x_{\Omega}$ the $x$ coordinate corresponding to the phase space point $\Omega$. The Fourier transforms of $p_{\rm s}$ for the two geometries of interest are
\begin{eqnarray}
  \tilde p_{\rm s}(\Omega,\Delta \omega) &=&
  \int_0^{\infty} d\tau p_{\rm s}(\Omega;\tau)
  e^{-i \tau \Delta \omega}
  \nonumber \\ 
  &=&
  \begin{cases}
    \frac{1}{2 (1 +  i \tau_{\rm D} \Delta \omega)} & (\mbox{CC}), \\
    \frac{\sinh[\pi (1-x_{\Omega}/L) \sqrt{i \tau_{\rm D} \Delta \omega}]}
    {\sinh[\pi \sqrt{i \tau_{\rm D} \Delta \omega}]}
  & (\mbox{WG}). \end{cases}~~~
\end{eqnarray}
Substituting Eq.\ (\ref{eq:Iavg}) for $\langle I \rangle$, we then find
\begin{eqnarray}
  C_{2,1} &=&
  \frac{\pi \hbar J_0(k \Delta y)^2}{2 p^2 W^2 T^2}
  \int d\Omega_B d\Omega_C
  |\tilde p_{\rm s}(\Omega_B;\Delta \omega)|^2
  p_{\rm d}(\Omega_C)^2
  \nonumber \\ && \mbox{} \times
  \frac{\partial}{\partial \tau_{\rm E}}
  p(\Omega_{C},\Omega_{B};\tau_{\rm E}).
\end{eqnarray}

In order to perform the integrations over the phase space points $\Omega_B$ and $\Omega_C$, we note that
\begin{equation}
  p(\Omega_C,\Omega_B;\tau) = \frac{1}{V_{\Omega}}
  e^{-\tau/\tau_{\rm D}}
  \label{eq:pchaotic}
\end{equation}
for a chaotic cavity, with $V_{\Omega}= 2 \pi A p/c = 4 p W \tau_{\rm D}$ the volume of the classical phase space. For a quasi-one-dimensional random waveguide, one has
\begin{eqnarray}
  p(\Omega_C,\Omega_B;\tau) &=&
  \frac{2}{V_{\Omega}}
  \sum_{n=1}^{\infty} \sin \frac{n \pi x_{C}}{L} \sin \frac{n \pi x_{B}}{L}
  e^{-n^2 \tau/\tau_{\rm D}}, \nonumber \\ 
  \label{eq:wg}
\end{eqnarray}
with phase space volume $V_{\Omega}= 2 \pi L W p/c$. In both cases, one finds that the contribution $C_{2,1}$ to the speckle correlation function has the form
\begin{eqnarray}
  C_{2,1} &=&
  \frac{1}{g}
  J_0(k \Delta y)^2 A_2(\Delta \omega),
  \label{eq:C21general}
\end{eqnarray}
where
\begin{equation}
  A_2(\Delta \omega) = 
  - \frac{e^{-\tau_{\rm E}/\tau_{\rm D}}}
  {4 (1 + (\tau_{\rm D} \Delta \omega)^2)}.
\end{equation}
for the case of a chaotic cavity and
\begin{eqnarray}
  A_{2}(\Delta \omega) &=& 2
  \sum_{n=1}^{\infty} (-1)^n
  e^{-n^2 \tau_{\rm E}/\tau_{\rm D}}
    \nonumber \\ && \mbox{} \times
  a_2(n,0) a_2(n, \sqrt{2 \pi^2 \tau_{\rm D} \Delta \omega})
\end{eqnarray}
for the case of the quasi-one-dimensional random waveguide, with
\begin{eqnarray}
  a_2(n,z) &=& \frac{n^2 \pi^2}{n^2 \pi^2 + z^2}
  \nonumber \\ && \mbox{} \times
  \left[ 1
  - \frac{2 z^2 (\cos z - (-1)^n)}{(n^2 \pi^2 - z^2)(\cosh z - \cos z)}
  \right].
\end{eqnarray}  
Similarly, one finds
\begin{eqnarray}
  C_{2,2} &=&
  \frac{1}{g}
  J_0(k \Delta Y)^2 A_2(\Delta \omega) e^{-\tau_{\rm E}/\tau_{\rm D}}.
\end{eqnarray}

\begin{figure}[tb]
\includegraphics[width=\columnwidth]{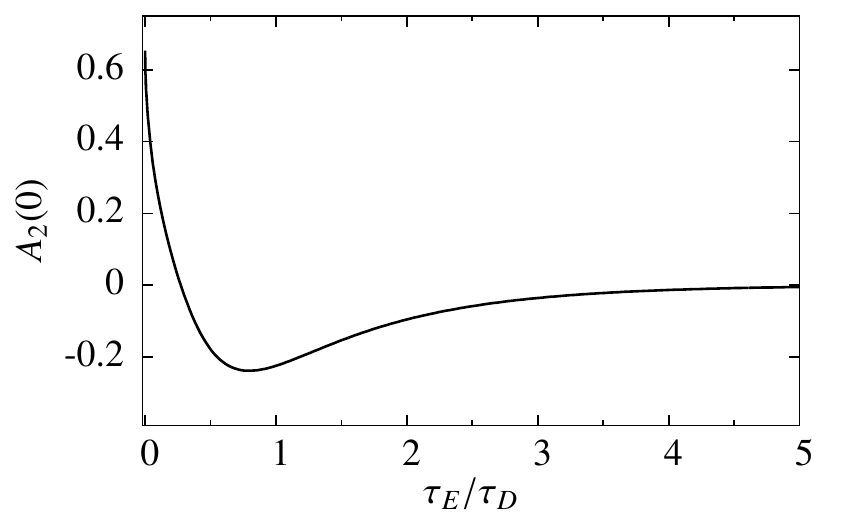}
\caption{The function $A_2(0)$ versus $\tau_{\rm E}/\tau_{\rm D}$ for the case of a quasi-one-dimensional random waveguide.\label{alpha}}
\end{figure}

The results of diagrammatic perturbation theory are reproduced in the limit $\tau_{\rm E} \to 0$, which gives
\begin{equation}
  A_2(\Delta \omega) =
  - \frac{1}{4(1 + (\tau_{\rm D} \Delta \omega)^2)}
\end{equation}
for a chaotic cavity and 
\begin{equation}
  A_2(\Delta \omega) = 
  2 a_2(\sqrt{2 \pi^2 \tau_{\rm D} \Delta \omega}),
\end{equation}
with
\begin{equation}
  a_2(z) = \frac{\sinh z - \sin z}{z(\cosh z - \cos z)}
\end{equation}
for the quasi-one-dimensional random waveguide.
The complete Ehrenfest-time dependence of the function $A_2(\Delta \omega)$ for a random waveguide and $\Delta \omega = 0$ is shown in Fig.\ \ref{alpha}.
  
\section{$C_{3}$ - contribution}
\label{sec:5}

\begin{figure*}
\includegraphics[width=\textwidth]{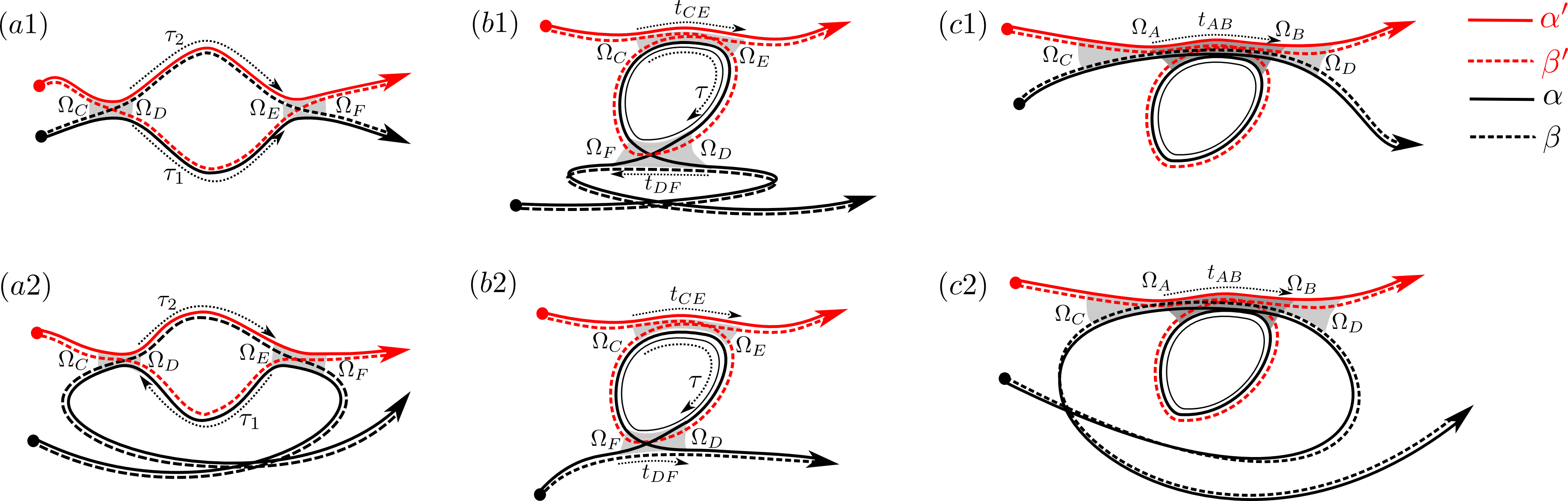}
\caption{(color online) Three different trajectory constellations contribution to $C_{3}$. All three constellations contain two small-angle encounters. The bright gray areas indicate single encounters, the dark grey areas indicate overlapping encounters. The constellations (b1), (b2), (c1) and (c2) involve a periodic reference trajectory $\gamma$, shown thin. The figure shows only the ``minimal'' version of the constellations (b1), (b2), (c1), and (c2), in which the trajectory $\alpha$ winds once around $\gamma$, and $\alpha'$ does not wind around $\gamma$ at all. Other contributions, in which $\alpha$ and $\alpha'$ wind $n$ and $n-1$ times around $\gamma$, respectively, with $n > 1$, are not shown in the figure, as well as constellations in which $\alpha$ and $\alpha'$ wind $n-1$ and $n$ times around $\gamma$, respectively, with $n \ge 1$. \label{2-enc}}
\end{figure*}

The $C_3$ contribution, which describes the correlation of intensities to leading order in $1/g$ if the sources and the detectors are a distance much longer than the wave length apart, has contributions from three different families of trajectory constellations, which are depicted in Fig. \ref{2-enc}. These families have in common, that the trajectories have two small-angle encounters, such that $\alpha$ is paired with $\beta$ and $\alpha'$ is paired with $\beta'$ before the first encounter and after the second encounter, whereas $\alpha$ is paired with $\beta'$ and $\alpha'$ with $\beta$ between the encounters. In constellations (a1) and (a2) the two encounters are traversed sequentially. In the constellation (b1) and (b2) the two encounters lie on the same periodic trajectory $\gamma$ and may or may not overlap. (The figure shows the non-overlapping case only.) In constellations (c1) and (c2) the two encounters overlap, too. However, unlike in configuration (b1) and (b2), only part of each 
encounter lies on a periodic trajectory. The constallations labeled (1) and (2) differ in the direction the two encounters are traversed: For the constellations (a1), (b1), and (c1), all trajectories pass through all encounters in the same direction, whereas for (a2), (b2), and (c2), the trajectories $\alpha$ and $\beta$ pass through the encounters in the opposite direction as the trajectories $\alpha'$ and $\beta'$.

Similar trajectory constellations were considered in the calculation of the conduction fluctuations \cite{brouwer2006,brouwer2007b}. The action difference $\Delta S$ consists of two contributions only,
\begin{equation}
  \Delta S = \Delta S_{\omega} + \Delta S_{\rm enc}.
\end{equation}
Since the action difference $\Delta S$ for the trajectories of Fig.\ \ref{2-enc} contain no contributions from the source or the detector, the calculation of the $C_3$ correlation function proceeds largely parallel to the calculation of the conductance fluctuations in Ref.\ \onlinecite{brouwer2007b}. We write the $C_3$ correlation function as a sum of six terms, $C_{3} = C_{3,a1} + C_{3,a2} + C_{3,b1} + C_{3,b2} + C_{3,c1} + C_{3,c2}$, and discuss each of those terms separately.

Following Ref.\ \onlinecite{brouwer2007b}, one finds that the contributions $
C_{3,a1}$ and $C_{3,a1}$ read
\begin{widetext}
\begin{eqnarray}
  C_{3,a1} &=&
  \left(\frac{\hbar^2 \pi^2}{16 p^2 W^2 \langle I \rangle}\right)^{2}
  \int d\Omega_C\Omega_D\Omega_E\Omega_F \int_0^{\infty} d\tau_1 d\tau_2
  p_{\rm d}(\Omega_F)^2 \left[\frac{\partial}{\partial\tau_{\rm E}}
  p(\Omega_F,\Omega_E;\tau_{\rm E})\right]
  \nonumber\\  && \mbox{} \times 
  p(\Omega_E,\Omega_D;\tau_1)
  p(\Omega_E,\Omega_D;\tau_2) 
  e^{i (\tau_1-\tau_2) \Delta \omega}
  \left[\frac{\partial}{\partial\tau_{\rm E}}
  p(\Omega_D,\Omega_C;\tau_{\rm E})\right]
  p_{\rm s}(\bar{\Omega}_{C})^2, \\
\label{eq:2-enc_a_1}
  C_{3,a2} &=&
  \left(\frac{\hbar^2 \pi^2}{16 p^2 W^2 \langle I \rangle}\right)^{2}
  \int d\Omega_C\Omega_D\Omega_E\Omega_F \int_0^{\infty} d\tau_1 d\tau_2
  p_{\rm d}(\Omega_F) p_{\rm s}(\bar{\Omega}_F)
  \left[\frac{\partial}{\partial\tau_{\rm E}}
  p(\Omega_F,\Omega_E;\tau_{\rm E})\right]
  \nonumber\\  && \mbox{} \times 
  p(\Omega_E,\Omega_D;\tau_1)
  p(\Omega_E,\Omega_D;\tau_2) 
  e^{i (\tau_1-\tau_2) \Delta \omega}
  \left[\frac{\partial}{\partial\tau_{\rm E}}
  p(\Omega_D,\Omega_C;\tau_{\rm E})\right]
  p_{\rm d}(\Omega_C) p_{\rm s}(\bar{\Omega}_{C}).
\end{eqnarray}
The definition of the phase space points $\Omega_C$, $\Omega_D$, $\Omega_E$, and $\Omega_F$, as well as the propagation times $\tau_1$ and $\tau_2$ is shown in Fig.\ \ref{2-enc}a1 and a2. Further, $p_{\rm s}(\Omega)$ is the probability that a trajector originating in the phase space point $\Omega$ exits the medium at the source side. Similarly, for $C_{3,b1}$ and $C_{3,b2}$ we find
\begin{eqnarray}
  \label{eq:2-enc_b_1} 
  C_{3,b1} &=&
  \left(\frac{\hbar^2 \pi^2}{16 p^2 W^2 \langle I \rangle}\right)^{2}
  \int d\Omega_C\Omega_D\Omega_E\Omega_F 
  \int_0^{\infty} d\tau_{\gamma} \int_0^{\tau_{\gamma}} d\tau 
  p_{\rm d}(\Omega_F) p_{\rm d}(\Omega_E)
  p_{\rm s}(\bar{\Omega}_D) p_{\rm s}(\bar{\Omega}_C)
  \nonumber\\  && \mbox{} \times 
  (e^{i \tau_{\gamma} \Delta \omega} + e^{-i \tau_{\gamma} \Delta \omega})
  \left[ \frac{\partial}{\partial t_{DF}} \frac{\partial}{\partial t_{CE}}
  p_{\gamma}(\Omega_C,\Omega_D,\Omega_E,\Omega_F;\tau_{\gamma},\tau,t_{CE},t_{DF}) \right]_{t_{CE} = t_{DF} = \tau_{\rm E}}. \\
  C_{3,b2} &=&
  \left(\frac{\hbar^2 \pi^2}{16 p^2 W^2 \langle I \rangle}\right)^{2}
  \int d\Omega_C\Omega_D\Omega_E\Omega_F 
  \int_0^{\infty} d\tau_{\gamma} \int_0^{\tau_{\gamma}} d\tau
  p_{\rm s}(\bar{\Omega}_F) p_{\rm d}(\Omega_E)
  p_{\rm d}(\Omega_D) p_{\rm s}(\bar{\Omega}_C)
  \nonumber\\  && \mbox{} \times 
  (e^{i \tau_{\gamma} \Delta \omega} + e^{-i \tau_{\gamma} \Delta \omega})
  \left[ \frac{\partial}{\partial t_{DF}} \frac{\partial}{\partial t_{CE}}
  p_{\gamma}(\Omega_C,\Omega_D,\Omega_E,\Omega_F;\tau_{\gamma},\tau,t_{CE},t_{DF}) \right]_{t_{CE} = t_{DF} = \tau_{\rm E}},
\end{eqnarray}
where the definition of the phase space points and time intervals is given in panels b1 and b2 of Fig.\ \ref{2-enc}. The function $p_{\gamma}(\Omega_C,\Omega_D,\Omega_E,\Omega_F;\tau_{\gamma},\tau,t_{CE},t_{DF})$ is the probability density for the phase space points $\Omega_C$, $\Omega_D$, $\Omega_E$, and $\Omega_F$ to lie on the same periodic trajectory $\gamma$ of period $\tau_{\gamma}$ with the specified time intervals for propagation between them. In order to describe the case that the trajectories $\alpha$ and/or $\alpha'$ wind multiple times around $\gamma$, the propagation times $t_{CE}$ and $t_{DF}$ are allowed to be larger than $\tau_{\gamma}$. Defining
\begin{eqnarray}
  \tau_{CE} = t_{CE}\text{ mod }\tau_{\gamma},\ \
  \tau_{DF} = t_{DF}\text{ mod }\tau_{\gamma},\ \
\label{eq:taus}
\end{eqnarray}
we can then write $p_{\gamma}(\Omega_C,\Omega_D,\Omega_E,\Omega_F;\tau_{\gamma},\tau,t_{CE},t_{DF})$ as \cite{brouwer2007b}
\begin{eqnarray}
  \lefteqn{p_{\gamma}(\Omega_C,\Omega_D,\Omega_E,\Omega_F;\tau_{\gamma},\tau,t_{CE},t_{DF})} \nonumber \\ &=&
  p(\Omega_{E},\Omega_{C};\tau_{CE})p(\Omega_{D},\Omega_{E};\tau_{}-\tau_{CE})p(\Omega_{F},\Omega_{D};\tau_{DF})p(\Omega_C,\Omega_{F};\tau_{\gamma}-\tau_{}-\tau_{DF})\nonumber \\
&& \mbox{} +p(\Omega_{D},\Omega_{C};\tau_{})p(\Omega_{E},\Omega_{D};\tau_{CE}-\tau_{})p(\Omega_{F},\Omega_{E};\tau_{DF}+\tau_{}-\tau_{CE})p(\Omega_C,\Omega_{F};\tau_{\gamma}-\tau_{}-\tau_{DF})\nonumber \\
&&  \mbox{} +p(\Omega_{F},\Omega_C;\tau_{}+\tau_{DF}-\tau_{\gamma})p(\Omega_{E},\Omega_{F};\tau_{CE}-\tau_{DF}-\tau_{}+\tau_{\gamma})p(\Omega_{D},\Omega_{E};\tau_{}-\tau_{CE})p(\Omega_C,\Omega_{D};\tau_{\gamma}-\tau_{})\nonumber \\
&&  \mbox{} +p(\Omega_{D},\Omega_C;\tau_{})p(\Omega_{F},\Omega_{D};\tau_{DF})p(\Omega_{E},\Omega_{F};\tau_{CE}-\tau_{DF}-\tau_{})p(\Omega_C,\Omega_{E};\tau_{\gamma}-\tau_{CE})\nonumber \\
&& \mbox{} +p(\Omega_{E},\Omega_C;\tau_{CE})p(\Omega_{F},\Omega_{E};\tau_{DF}+\tau_{}-\tau_{CE}-\tau_{\gamma})p(\Omega_{D},\Omega_{F};\tau_{\gamma}-\tau_{DF})p(\Omega_C,\Omega_{D};\tau_{\gamma}-\tau_{})\nonumber \\
&& \mbox{} +p(\Omega_{F},\Omega_C;\tau_{}+\tau_{DF}-\tau_{\gamma})p(\Omega_{D},\Omega_{F};\tau_{\gamma}-\tau_{DF})p(\Omega_{E},\Omega_{D},\tau_{CE}-\tau_{})p(\Omega_C,\Omega_{E};\tau_{\gamma}-\tau_{CE}),
\label{eq:P_b}
\end{eqnarray}
where we use the convention that the probability density $p(\Omega,\Omega';\tau) = 0$ for $\tau < 0$.
Finally, the constellations (c1) and (c2) are corrections to (b1) and (b2), which take into account the influence of correlated propagation between $\alpha$ and $\alpha'$ before and after encounter with periodic trajectory $\gamma$. For these constellations, one finds
\begin{eqnarray}
  C_{3,c1} &=&
  \left( \frac{\pi^{2}}{16\, k^{2}W^{2} \langle I \rangle}\right)^{2}
  \int d\Omega_C\Omega_A\Omega_B\Omega_D \int d\tau_{\gamma}
  \int_0^{\tau_{\rm E}} dt_{AB}
  p_{\rm d}(\Omega_D)^2
  \left[\partial_{\tau_{\rm E}} p(\Omega_A,\Omega_C,\tau_{\rm E}-t_{AB})\right]   \nonumber \\ && \mbox{} \times 
  p_{\gamma}(\Omega_{A},\Omega_{B};\tau_{\gamma},t_{AB})
  \left[\partial_{\tau_{\rm E}} p(\Omega_D,\Omega_B,\tau_{\rm E}-t_{AB})\right]
  (e^{i \tau_{\gamma} \Delta \omega} + e^{-i \tau_{\gamma} \Delta \omega})
  p_{\rm s}(\bar{\Omega}_C)^2, 
\label{eq:2-enc_c_2} \\
  C_{3,c2} &=&
  \left( \frac{\pi^{2}}{16\, k^{2}W^{2} \langle I \rangle}\right)^{2}
  \int d\Omega_C\Omega_A\Omega_B\Omega_D \int d\tau_{\gamma}
  \int_0^{\tau_{\rm E}} dt_{AB}
  p_{\rm d}(\Omega_D)
  p_{\rm s}(\bar{\Omega}_D)
  \left[\partial_{\tau_{\rm E}} p(\Omega_A,\Omega_C,\tau_{\rm E}-t_{AB})\right]   \nonumber \\ && \mbox{} \times 
  p_{\gamma}(\Omega_{A},\Omega_{B};\tau_{\gamma},t_{AB})
  \left[\partial_{\tau_{\rm E}} p(\Omega_D,\Omega_B,\tau_{\rm E}-t_{AB})\right]
  (e^{i \tau_{\gamma} \Delta \omega} + e^{-i \tau_{\gamma} \Delta \omega})
  p_{\rm s}(\bar{\Omega}_C) 
  p_{\rm d}(\Omega_C), 
\end{eqnarray}
where $p_{\gamma}(\Omega_{A},\Omega_{B};\tau_{\gamma},t_{AB})$ is the probability density that the phase space points $\Omega_{A}$ and $\Omega_{B}$ lie on one periodic trajectory $\gamma$ with period $\tau_{\gamma}$ and the propagation time $t_{AB}$ between them as indicated in Fig.\ \ref{2-enc}c1 and c2. One has
\begin{equation}
  p_{\gamma}(\Omega_{A},\Omega_{B};\tau_{\gamma},t_{AB})
  = p(\Omega_{A},\Omega_{B};\tau_{AB}) p(\Omega_{B},\Omega_{A};\tau_{\gamma} - \tau_{AB}),
\end{equation}
\end{widetext}
with 
\begin{equation}
  \tau_{AB} = t_{AB}\text{ mod }\tau_{\gamma}.
\end{equation}

We now proceed with the calculation of $C_3$ for the cases of a chaotic cavity and a quasi-one-dimensional random waveguide separately.

\subsection{chaotic cavity}

For the chaotic cavity we insert the known expressions for the classical propagators and probabilities, $p_{\rm s}(\Omega) = p_{\rm d}(\Omega) = 1/2$ and $p(\Omega,\Omega';\tau) = \Omega^{-1} e^{-\tau/\tau_{\rm D}}$, see Eqs.\ (\ref{eq:pchaotic1}) and (\ref{eq:pchaotic}), and find
\begin{eqnarray}
  C_{3,a1} &=& C_{3,a2} \nonumber \\ &=&
  \frac{ e^{-2 \tau_{\rm E}/\tau_{\rm D}}}{16 g^2 (1 + (\tau_{\rm D} \Delta \omega)^2)}, \\
  C_{3,b1} &=& C_{3,b2} \nonumber \\ &=& 0, \\
  C_{3,c1} &=& C_{3,c2} \nonumber \\ &=&
  \frac{1 - e^{-2 \tau_{\rm E}/\tau_{\rm D}}}{16 g^2 (1 + (\tau_{\rm D} \Delta \omega)^2)},
\end{eqnarray}
so that one arrives at the remarkably simple result
\begin{equation}
  C_3 = \frac{1}{g^2} A_3(\Delta \omega),
\end{equation}
with
\begin{equation}
  A_3(\Delta \omega) = \frac{1}{8(1 + (\tau_{\rm D} \Delta \omega)^2)}, 
\label{eq:2-enc_CC_result}
\end{equation}
independent of the ratio $\tau_{\rm E}/\tau_{\rm D}$. This observation is consistent with the observation that the conductance autocorrelation function is independent of $\tau_{\rm E}/\tau_{\rm D}$ \cite{brouwer2007}.

\subsection{random waveguide}

\begin{figure}[tb]
\includegraphics[width=\columnwidth]{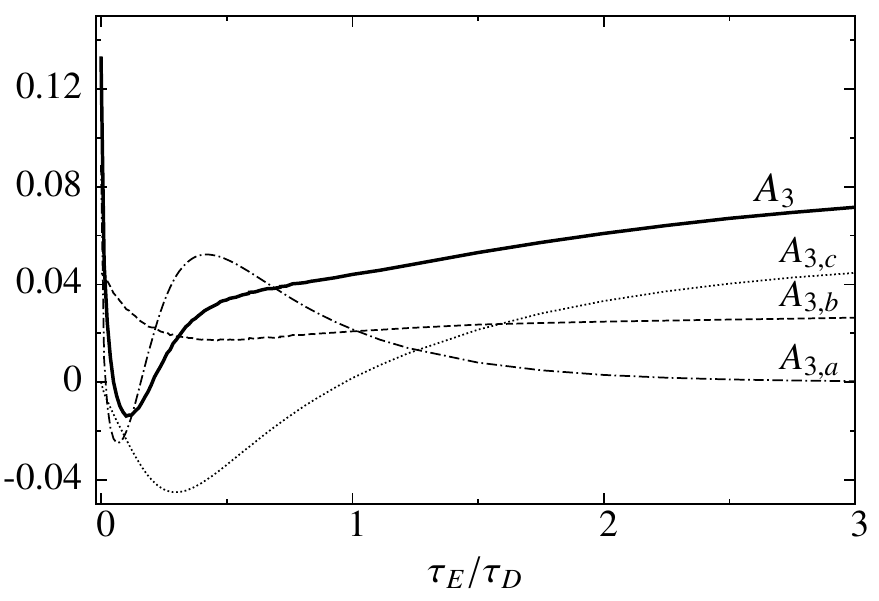}
\caption{The rescaled correlation function $A_3 = C_3 g^2$ at equal frequencies, $\Delta \omega = 0$, for the random waveguide, as a function of the Ehrenfest time $\tau_{\rm E}$, together with its three contributions $A_{3,a}$, $A_{3,b}$, and $A_{3,c}$. \label{beta}}
\end{figure}

The phase space probability densities $p_{\rm d}(\Omega)$ and $p(\Omega,\Omega';\tau)$ are given in Eqs.\ (\ref{eq:wg1}) and (\ref{eq:wg}), whereas $p_{\rm s}(\Omega) = 1-x_{\Omega}/L$. Instead of the correlation function $C_3(\Delta \omega)$ we calculate the ``form factor'',
\begin{equation}
  K_3(t) = \frac{1}{2 \pi} \int d\Delta \omega C_3(\Delta \omega) e^{i t \Delta \omega}.
\end{equation}
The calculation of the three contributions $K_{3,a2}$, $K_{3,b2}$ and $K_{3,c2}$ is identical to that of Ref.\ \onlinecite{brouwer2007b}. (In Ref.\ \onlinecite{brouwer2007b} these three contributions are called $K^{(a)}$, $K^{(b)}$, and $K^{(c)}$, respectively.) The calculation of $K_{3,a1}$, $K_{3,b1}$ and $K_{3,c1}$ differs with respect to details. The final results for the sums $K_{3,a} = K_{3,a1} + K_{3,a2}$, $K_{3,b} = K_{3,b1} + K_{3,b2}$, and $K_{3,c} = K_{3,c1} + K_{3,c2}$ read
\begin{eqnarray}
  K_{3,a}(t) &=& \frac{1}{g^{2}\tau_D}\sum_{\mu,\nu,\rho,\sigma}d_{\nu\mu\sigma}d_{\rho\mu\sigma}g_{\nu\rho}\nonumber \\
  && \times \frac{e^{-\sigma^{2}|t|/\tau_D}}{\mu^{2}+\sigma^{2}}  e^{-(\nu^{2}+\rho^{2})\tau_{E}/\tau_D},  \\
  K_{3,b}(t)&=&-\frac{2}{g^2\tau_{D}}\sum_{\mu,\nu,\rho,\sigma} c_{\mu\sigma}c_{\rho\nu}c_{\mu\nu}c_{\rho\sigma} \nonumber \\
  &&\left[\left(\mu^{2}-\nu^{2}\right)\left(\sigma^{2}-\rho^{2}\right)e^{-(\rho^{2}+\mu^{2})(|t|-|2\tilde{t}-|t||)/2\tau_{D}}\right. \nonumber \\
  &&\left.\times f_{\nu^{2},\sigma^{2}}\left(\frac{|2\tilde{t}-|t||}{\tau_{D}}\right)+2\left(\sigma^{2}-\mu^{2}\right)\left(\nu^{2}-\mu^{2}\right)\right.\nonumber \\
  &&\left.\times e^{-\nu^{2}|2\tilde{t}-|t||/\tau_{D}}f_{\sigma^{2}+\nu^{2},\rho^{2}+\mu^{2}}\left(\frac{|t|-|2\tilde{t}-|t||}{2\tau_{D}}\right)\right]\nonumber \\
  &&+\frac{2}{g^2\tau_{D}}\sum_{\rho}e^{-\rho^{2}|t|/\tau_{D}}\left(\frac{1}{6\pi^{2}}-\frac{1}{\rho^{2}\pi^{4}}\right), \\
  K_{3,c}(t) &=& \frac{1}{g^{2}\tau_D}\sum_{\mu,\nu,\rho,\sigma}d_{\nu\mu\sigma}d_{\rho\mu\sigma} g_{\nu\rho} \left[ e^{-\mu^{2}(|t|-\tilde{t})/\tau_D}\right. \nonumber  \\
  &&\times f_{\nu^{2}+\rho^{2}+\mu^{2},\sigma^{2}}\left(\frac{\tilde{t}}{\tau_D}\right)+f_{\nu^{2}+\rho^{2}+\mu^{2},\sigma^{2}}\left(\frac{|t|}{\tau_D}\right) \nonumber\\
  &&\left.\times\frac{e^{-(\nu^{2}+\rho^{2})\tilde{t}/\tau_D}-e^{-(\nu^{2}+\rho^{2})\tau_{E}/\tau_D}}{1-e^{-(\nu^{2}+\rho^{2})|t|/\tau_D}}\right],
\end{eqnarray}
where we abbreviated
\begin{equation}
  \tilde{t}=\tau_E\text{mod}|t|,
\end{equation}
\begin{equation}
  c_{\mu\nu} = \begin{cases}
  \frac{8\mu\nu}{\pi^2(\mu^2-\nu^2)^2} 
  & \mbox{if $\mu + \nu$ odd}, \\
  0 & \mbox{else}, \end{cases}
\end{equation}
\begin{equation}
  d_{\mu\nu\rho} = \begin{cases}
  \frac{16}{\pi^4} \sum_{\pm} \frac{\pm1}{\mu^2 - (\nu \pm \rho)^2}
  & \mbox{if $\mu + \nu + \rho$ odd}, \\
  0 & \mbox{else}, \end{cases}
\end{equation}
\begin{equation}
  g_{\mu\nu}=\begin{cases}
  2-\frac{\pi^{2}\mu^{2}}{2}+\frac{\pi^{4}\mu^{2}\nu^{2}}{16} & \mbox{if $\mu$ and $\nu$ odd}\\
  -(-1)^{\nu}\frac{\pi^{4}\mu^{2}\nu^{2}}{16} & \mbox{else}, \end{cases}
\end{equation}
\begin{equation}
  f_{\alpha,\beta}(x)=\begin{cases}
  xe^{-\beta x} & \mbox{if $\alpha=\beta$} \\
  \frac{e^{-\beta x}-e^{-\alpha x}}{\alpha-\beta} & \mbox{else}.
\end{cases} 
\end{equation}
The correlation functions can be obtained by Fourier transformation, $C_{3,i}=\int dtK_{3,i}e^{-it\Delta\omega}$. The full expressions are too lengthy to report here, which is why we restrict ourselves to the contributions at $\Delta \omega=0$, 
\begin{widetext}
\begin{eqnarray}
  \left. C_{3,a}\right|_{\Delta\omega=0}&=&\frac{2}{g^{2}}\sum_{\mu,\sigma}\frac{1}{\sigma^2(\mu^{2}+\sigma^{2})}
  \sum_{\nu,\rho}d_{\nu\mu\sigma}d_{\rho\mu\sigma}g_{\nu\rho}e^{-(\nu^{2}+\rho^{2})\tau_{E}/\tau_D}, \\ 
  \left. C_{3,b}\right|_{\Delta\omega=0}&=&-\frac{4}{g^2}\sum_{\mu,\nu,\rho,\sigma}c_{\mu\sigma}c_{\mu\nu}c_{\rho\nu}c_{\rho\sigma}\left\{ (\mu^{2}-\nu^{2})(\sigma^{2}-\rho^{2})
  \left[\frac{1}{\nu^{2}\sigma^{2}}e^{-(\mu^{2}+\rho^{2})\tau_{E}/\tau_{D}}
  +h_{\mu^{2}+\rho^{2},\sigma^{2},\mu^{2}+\rho^{2},\nu^{2}}^{(2)}\left(\frac{\tau_{E}}{\tau_{D}}\right)\right] \right.\nonumber\\
  &&\left.+2\left(\nu^{2}-\mu^{2}\right)\left(\sigma^{2}-\mu^{2}\right)
  \left[\frac{1}{\sigma^{2}}f_{\mu^{2}+\rho^{2},\nu^{2}+\sigma^{2}}\left(\frac{\tau_{E}}{\tau_{D}}\right)
  +h_{\mu^{2}+\rho^{2},\sigma^{2},\nu^{2}+\sigma^{2},\sigma^{2}}^{(1)}\left(\frac{\tau_{E}}{\tau_{D}}\right)\right]\right\}+\frac{1}{15g^2}\\
  \left. C_{3,c}\right|_{\Delta\omega=0}&=&\frac{1}{g^{2}}\sum_{\mu,\sigma}\sum_{\nu,\rho}g_{\nu\rho}d_{\nu\mu\sigma}d_{\rho\mu\sigma}\nonumber\\
  &&\times\left[\sum_{j}\frac{2}{j(j+1)}\frac{f_{\mu^{2}/j,\nu^{2}+\rho^{2}}\left(\frac{\tau_{E}}{\tau_{D}}\right)-f_{\sigma^{2}/(j+1),\nu^{2}+\rho^{2}}\left(\frac{\tau_{E}}{\tau_{D}}\right)}{\sigma^{2}/(j+1)-\mu^{2}/j}
  +\frac{1}{\mu^{2}}f_{\sigma^{2},\nu^{2}+\rho^{2}}\left(\frac{\tau_{E}}{\tau_{D}}\right)\right],
\end{eqnarray}
with
\begin{eqnarray}
h_{\alpha_{1}\alpha_{2}\beta_{1}\beta_{2}}^{(1,2)}\left(\frac{\tau_{E}}{\tau_{D}}\right) &=&\sum_{n}\sum_{\pm}\frac{1}{n(2n\mp1)}
\times\begin{cases}
-\frac{\partial}{\partial\alpha_{1,2}}f_{\frac{\alpha_{1}}{2n\mp1},\frac{\alpha_{2}}{n}}\left(\frac{\tau_{E}}{\tau_{D}}\right) & \mbox{ if }\beta_{1,2}=\alpha_{1,2}, \\
\frac{f_{\frac{\alpha_{1}}{2n\mp1},\frac{\alpha_{2}}{n}}\left(\frac{\tau_{E}}{\tau_{D}}\right)-f_{\frac{\beta_{1}}{2n\mp1},\frac{\beta_{2}}{n}}\left(\frac{\tau_{E}}{\tau_{D}}\right)}{\beta_{1,2}-\alpha_{1,2}} & \mbox{ else}.
\end{cases} 
\end{eqnarray}
\end{widetext}
Dividing out the common prefactor $1/g^2$, one finds the Ehrenfest-time dependence of the correlation function $A_3(\Delta \omega=0)$ of Eq.\ (\ref{eq:intro01}).
The Ehrenfest-time dependence of the three contributions to $A_3\equiv A_3(\Delta \omega=0)$ is shown in Fig.\ \ref{beta}. Remarkably, $A_3$ depends on the Ehrenfest time, but does not disappear in the limit $\tau_{\rm E}/\tau_{\rm D} \to \infty$. The same behavior was found for the conductance fluctuations in a random waveguide \cite{brouwer2007b}.

The results of the diagrammatic perturbation theory are reproduced in the limit $\tau_E\rightarrow0$. At $\Delta\omega=0$ we find 
\begin{eqnarray}
  A_3&=& A_{3,a}+A_{3,b}+A_{3,c} \nonumber \\
 &=&
  \frac{2}{15},
\end{eqnarray}
in agreement with Ref.\ \onlinecite{sebbah2002}.

\section{Conclusion}
\label{sec:6}

The semiclassical calculations of this article have shown that there is a consistent ray-optics-based picture for the three contributions $C_1$, $C_2$, and $C_3$ to the speckle correlation function. The ray-optics-based calculation links the distinctive spatial dependence of each of the three contributions --- doubly short range, mixed short range/long range, and doubly long range --- to a distinctive dependence on the Ehrenfest time $\tau_{\rm E}$, the threshold time at the crossover between the  ray- and wave-like propagation of a minimal wave packet: The short-range contribution $C_1$ is independent of $\tau_{\rm E}$, the mixed-range contribution $C_2$ vanishes in the limit of large $\tau_{\rm E}$, whereas the long-range contribution $C_3$ remains finite in the limit of large $\tau_{\rm E}$. 

Perhaps the latter observation is the most striking one: the long-range correlations described by $C_3$ are an unambiguous interference phenomenon, which continues to exist in the ray limit ({\em i.e.}, in the limit where a generic minimal wave packet follows a single ray). The origin of this remarkable effect is the same as the persistence of mesoscopic fluctuations of the electronic conductance in the classical limit \cite{brouwer2006}. In both cases, the effect arises from ray trajectories which are trapped near periodic rays internal to the random medium, thus extending their dwell time long enough that their dynamics becomes effectively wavelike.

In the limit of zero Ehrenfest time, our ray-based results agree with those obtained within diagrammatic perturbation theory, an intrinsically wave-based approach. 

Ehrenfest-time related phenomena have been originally predicted by Larkin and Ovchinnikov in the context of mesoscopic superconductivity \cite{larkin1968}. In the last two decades, manifestations in mesoscopic electronic transport have been investigated vigorously in the theory community \cite{aleiner1996,agam2000,adagideli2003,rahav2005,whitney2006,tworzydlo2004,jacquod2004,brouwer2006,brouwer2007b,petitjean2009,waltner2011,waltner2012}. At the same time, there has been remarkably little experimental activity \cite{yevtushenko2000,oberholzer2002}. Main problems are the difficulty to obtain the required high-mobility samples and the impossibility to significantly vary the relevant time scales $\tau_{\rm E}$ and $\tau_{\rm D}$ without affecting the underlying classical dynamics \cite{oberholzer2002}. In order to circumvent this problem, Ref.\ \onlinecite{yevtushenko2000} considers the competition of $\tau_{\rm E}$ and the temperature-dependent dephasing time, thereby having to deal with a large theoretical 
and 
experimental uncertainty of the latter \cite{altland2007}.

Against this background, the purpose of the present calculation is to proceed towards the possibility that Ehrenfest-time related phenomena can be observed using optical or microwave techniques. Especially in the context of microwave experiments, the almost complete control over sample geometry facilitates a quantitative comparison with theory (see, {\em e.g.}, Refs.\ \onlinecite{dembowski2000,kim2005,dietz2007,hemmady2005,bellec2013} for a number of recent reports). The possibility to measure and analyze the speckle correlation function and its three contributions $C_1$, $C_2$, and $C_3$ has been proven\cite{sebbah2002,stoeckmann1990,sridhar1991,graef1992,stein1992}. We hope that the availability of theoretical predictions for the Ehrenfest-time dependence of these three copmonents will stimulate further experiments in this direction.

\acknowledgements

Financial support was granted by the Alexander von Humboldt Foundation in the framework of the Alexander von Humboldt Professorship, endowed by the Federal Ministry of Education and Research (PWB). Financial support by the Deutsche Forschungsgemeinschaft through Research Training Group GRK 1621 is gratefully acknowledged (MB).

\bibliography{refs}

\end{document}